\documentclass[sigconf]{acmart}
\AtBeginDocument{%
  \providecommand\BibTeX{{%
    \normalfont B\kern-0.5em{\scshape i\kern-0.25em b}\kern-0.8em\TeX}}}

\newcommand{\proj}{{\tt GWLZ}}

\usepackage{color}
\usepackage{svg}
\usepackage{soul}
\usepackage{placeins} 
\usepackage{multirow} 
\usepackage{url} 
\usepackage[misc]{ifsym}

\copyrightyear{2024}
\acmYear{2024}
\setcopyright{rightsretained}
\acmConference[FlexScience'24]{Workshop on AI and Scientific Computing at Scale using Flexible Computing Infrastructures}{June 3--4, 2024}{Pisa, Italy}
\acmBooktitle{Workshop on AI and Scientific Computing at Scale using Flexible Computing Infrastructures (FlexScience'24), June 3--4, 2024, Pisa, Italy}\acmDOI{10.1145/3659995.3660041}
\acmISBN{979-8-4007-0642-4/24/06}
\begin{document}
\title{GWLZ: A Group-wise Learning-based Lossy Compression Framework for Scientific Data}
\author{Wenqi Jia}
\email{wxj1489@mavs.uta.edu}
\affiliation{%
  \institution{University of Texas at Arlington}
  \city{Arlington}
  \state{TX}
  \country{USA}
  \postcode{76019}
}
\author{Sian Jin}
\email{sian.jin@temple.edu}
\affiliation{%
  \institution{Temple University}
  \city{Philadelphia}
  \state{PA}
  \country{USA}
  \postcode{19122}
}
\author{Jinzhen Wang}
\email{jinzhen.wang@brooklyn.cuny.edu}
\affiliation{%
  \institution{Brooklyn College, CUNY}
  \city{Brooklyn}
  \state{NY}
  \country{USA}
  \postcode{11210}
}
\author{Wei Niu}
\email{wniu@uga.edu}
\affiliation{%
  \institution{University of Georgia}
  \city{Athens}
  \state{GA}
  \country{USA}
  \postcode{30602}
}

\author{Dingwen Tao}
\email{ditao@iu.edu}
\affiliation{%
  \institution{Indiana University}
  \city{Bloomington}
  \state{IN}
  \country{USA}
  \postcode{47405}
}
\author{Miao Yin$^\dagger$}
\email{miao.yin@uta.edu}
\affiliation{%
  \institution{University of Texas at Arlington}
  \city{Arlington}
  \state{TX}
  \country{USA}
  \postcode{76019}
}

\renewcommand{\shortauthors}{Jia and Yin, et al.}
\begin{abstract}
The rapid expansion of computational capabilities and the ever-growing scale of modern HPC systems present formidable challenges in managing exascale scientific data. Faced with such vast datasets, traditional lossless compression techniques prove insufficient in reducing data size to a manageable level while preserving all information intact. In response, researchers have turned to error-bounded lossy compression methods, which offer a balance between data size reduction and information retention.
However, despite their utility, these compressors employing conventional techniques struggle with limited reconstruction quality. To address this issue, we draw inspiration from recent advancements in deep learning and propose \proj, a novel group-wise learning-based lossy compression framework with multiple lightweight learnable enhancer models. Leveraging a group of neural networks, \proj~ significantly enhances the decompressed data reconstruction quality with negligible impact on the compression efficiency.
Experimental results on different fields from the Nyx dataset demonstrate remarkable improvements by \proj, achieving up to 20\% quality enhancements with negligible overhead as low as 0.0003$\times$. \let\thefootnote\relax\footnotetext{$^\dagger$ Corresponding author.}
\end{abstract}



\maketitle

\section{Introduction} 
The rapid growth of computational power has facilitated the execution of complex scientific simulations across various fields of science. Users harness supercomputers to conduct these simulations and extract insights from the resulting data. However, despite the acceleration of simulations provided by supercomputers, users often encounter limitations in data storage and internet bandwidth on their end, as they may need to analyze the data locally, and some users also need to distribute large volumes of data across multiple endpoints via a data-sharing web service. The immense volume and rapid flow of data typically encountered pose significant challenges in terms of storage and transmission, highlighting a pressing necessity for efficient data compression techniques.

At the outset, researchers develop lossless compressors, e.g., LZ77 \cite{ziv1977universal}, GZIP \cite{peter1996gzip}, FPZIP \cite{lindstrom2006fast}, Zlib \cite{gailly2004zlib}, and Zstandard \cite{collet2015zstandard}, to address this issue. However, lossless compression techniques struggle to achieve significant compression ratios (typically 1$\times$-3$\times$) when applied to scientific data \cite{zhao2020sdrbench}. This limitation arises because lossless compression methods rely on repeated byte-stream patterns, whereas scientific data commonly consists of diverse floating-point numbers. Recently, the researchers proposed error-bounded lossy compression as a powerful method to address this issue. 

Compared to lossless compressors, lossy compressors \cite{lindstrom2014fixed, di2016fast, liu2022dynamic, liu2023high, zhao2021optimizing, tao2017significantly, liang2021mgard+, liang2019significantly, tao2019optimizing, zhao2020significantly, liang2018error} can achieve significantly higher compression ratios (e.g., 3.3$\times$ to 436$\times$ for SZ \cite{di2016fast}) while retaining essential information based on user-specified error bounds. Among these lossy compressors, many utilize predictive methods such as curve-fitting \cite{di2016fast} or spline interpolation \cite{lakshminarasimhan2013isabela} to anticipate the data. The compression ratio primarily hinges on the local smoothness of the data. However, given the inherent complexity of scientific simulations, spiky data points are often present within the dataset. Consequently, predictive-based compressors typically necessitate storing true values for these spiky data points, resulting in a diminished compression ratio while upholding the same level of data distortion. To build effective predictors that fully account for data variability, researchers have explored various techniques. For instance, Zhao et al. introduces higher-order predictors, specifically a 2nd-order predictor \cite{zhao2020significantly}, to enhance performance. Additionally, Tao et al. focuses on improving prediction accuracy for spiky data by employing multidimensional predictors \cite{tao2017significantly}.
Additionally, employing a selection mechanism from a pool of predictors can be beneficial \cite{di2016fast, tao2017significantly}. Di et al. identifies the optimal predictor with just two bits during compression, and the predictor size can be negligible, thus enhancing the compression ratio \cite{di2016fast}. Meanwhile, Tao et al. introduces a range of multilayer prediction formulas for users to choose from, as different datasets may exhibit varying degrees of effectiveness with specific layer values for prediction \cite{tao2017significantly}. These predictor-based compressors predominantly employ adaptive error-controlled quantization to achieve controlled error levels. Even though existing lossy compression algorithms can achieve a very high compression ratio, there is still a significant gap between the decompressed data and the original scientific data, especially under a high error bound, thereby limiting the reconstruction quality and hindering scientific discovery.

In the past decades, deep neural networks (DNNs) have found propelled breakthroughs across a spectrum of foundational and applied tasks in artificial intelligence, including but not limited to image classification \cite{rawat2017deep}, object detection \cite{liu2020deep}, action recognition \cite{kong2022human}, super-resolution \cite{wang2020deep}, and natural language processing \cite{torfi2020natural}. Researchers attempt to leverage DNN models, e.g., \cite{liu2021exploring, liu2023srn}, to improve the compression quality. However, incorporating DNNs to lossy compression algorithms is non-trivial, as the main challenges have still not been addressed: \ul{1)} to remember important information successfully, the DNN model size is generally large, e.g., the HAT model has approximately 9M parameters \cite{liu2023srn}, and the AE model has around 1M parameters \cite{liu2021exploring}. Hence, it inevitably introduces significant storage overhead as we save these models to enhance lossy compression. \ul{2)} DNN models often require retraining when applied to data from different scientific domains, which are very costly in terms of time and resources. Alternatively, developers may provide pre-trained weights for the models to load, eliminating the need for per-input training. However, this approach may lack specificity, potentially compromising performance guarantees. \ul{3)} DNN models may incur extra computational costs compared to traditional methods, such as interpolation-based predictors, which typically have linear time complexity.

Our paper aims to mitigate these challenges and significantly improve existing lossy compression quality. We introduce \proj, a group-wise learning-based lossy compression framework. The key innovation of our \proj~ lies in \ul{\textit{learning the residual information between the decompressed data generated by lossy compressors and the original data in groups with multiple lightweight DNN models based on encoder-decoder architecture}}. In \proj, the data will be partitioned into multiple small sub-groups, and each lightweight model is responsible for enhancing one associated sub-group. In other words, they serve as learned enhancers that can be appended to lossy compressors to boost the quality of decompressed data. Moreover, thanks to lightweight nature, these DNN models offer the advantage of negligible storage and computational complexity, as well as short training and inference times. Additionally, unlike compressors that require pre-training \cite{liu2021exploring, liu2023srn}, \proj~ enjoys a very short learning time to train these lightweight models fully. More importantly, by distributing the workload among multiple models, the overall compression process becomes more scalable and adaptable to diverse scientific data. In summary, the contributions of the paper are itemized as follows:
\begin{itemize}
\item We propose \proj, a novel learning-based lossy compression framework, where multiple DNN models are considered lightweight enhancers trained simultaneously with the lossy compression process and attached to the compressed data with negligible cost. To the best of our knowledge, our \proj, for the first time, addresses the aforementioned issues as a learn-to-compress framework.

\item By leveraging a comprehensive analysis of the scientific data during compression and reconstruction, we developed a group-wise learning methodology to empower our lightweight DNN-based enhancers. Specifically, the group-wise strategy divides data into multiple groups to achieve a well-balanced data distribution and magnitude ranges. Then, each group is assigned a single lightweight model, enabling a smooth learning process to mitigate the reconstruction error within the group well. This approach ensures that each model focuses on a specific sub-group of the data, enhancing its ability to capture nuanced variations within that sub-group. 

\item We carry out numerous experiments to assess the efficacy of our approach. The experiment results demonstrate a significant improvement in reconstruction quality under the same compression ratio level. Notably, experiments on the Nyx Cosmological Simulation Dataset indicate a potential improvement of reconstruction quality by up to 20\%, with negligible impact on the compression ratio.
\end{itemize}

\section{Background and Motivation}

\begin{figure*}[t!]
  \centering
  \begin{minipage}[t]{0.49\textwidth}
    \includegraphics[width=\linewidth]{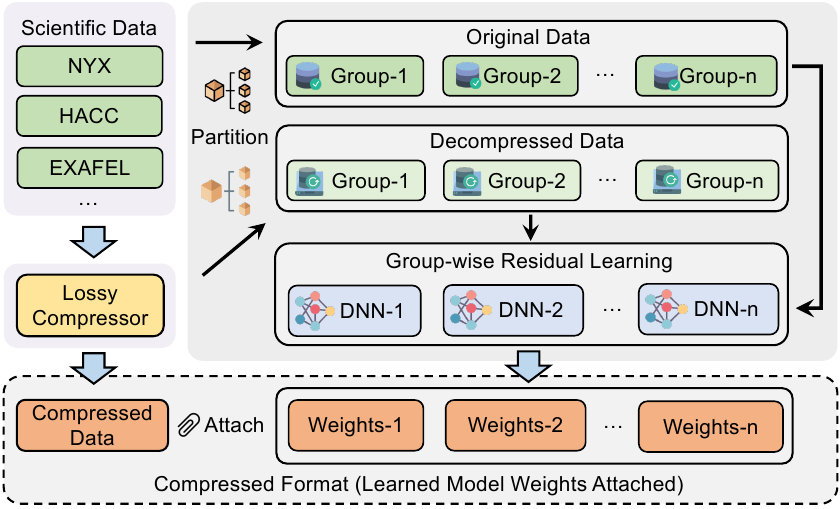}
    \caption{Overview of the \proj~compression module.}
    \label{fig:framework-compr}
  \end{minipage}
  \hfill 
  \begin{minipage}[t]{0.49\textwidth}
    \includegraphics[width=\linewidth]{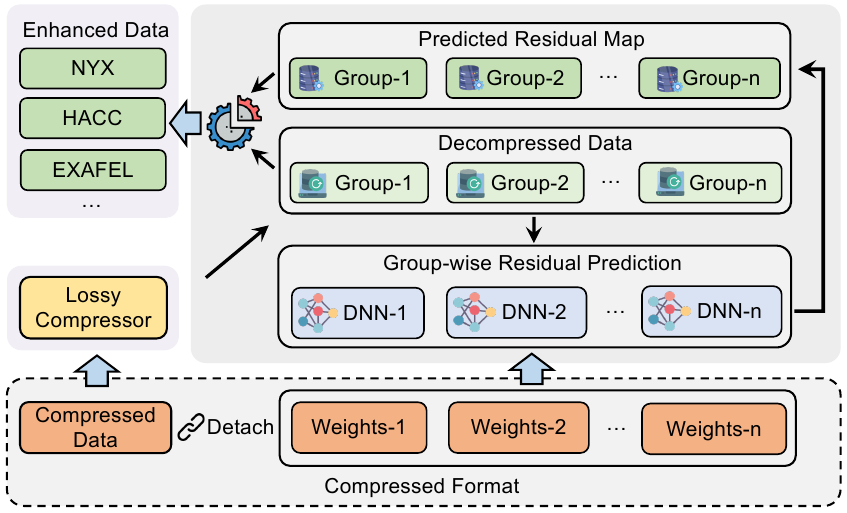}
    \caption{Overview of the \proj~reconstruction module. }
    \label{fig:framework-decompr}
  \end{minipage}
\end{figure*}

\subsection{Lossy Compression}
\label{section_lossy-compression}
Lossy compression is a widely used technique for reducing data size, accomplished by sacrificing certain non-essential information, thereby enabling high compression ratios. With an increasing number of research efforts in this domain, the field of lossy compression, particularly prediction-based compression, is expanding \cite{zhao2020sdrbench}. Researchers have progressively refined lossy compression frameworks \cite{tao2019optimizing, liang2022sz3, tao2019z, liu2022dynamic, liu2023faz, liu2023high}. Tao et al. begins by creating a selection method to optimize between SZ and ZFP compressors \cite{tao2019optimizing}. Liang et al. further advances customization with the modular SZ3 framework \cite{liang2022sz3}. Data analysis and evaluation are streamlined with Tao et al.'s Z-checker \cite{tao2019z}. Liu et al. introduces auto-tuning with QoZ, incorporating quality metrics and dynamic dimension freezing \cite{liu2022dynamic}. Their FAZ framework builds upon these advancements, offering a wider array of data transforms and predictors for even greater pipeline customization \cite{liu2023faz}.

Error-bounded lossy compression offers users the flexibility to select their desired error bound, denoted by $e$. For the given error bound $e$, different lossy compressors may exhibit varying compression quality. Within the lossy compression community, two primary metrics are typically employed to evaluate the compression quality of a lossy compressor: compression ratio and data distortion.

\textbf{Compression Ratio.} The compression ratio $CR$ can be defined as the ratio of the size of the input data $|X|$ to the size of the compressed data $|Z|$. Mathematically, it can be expressed as $CR=|X|/|Z|$. It provides a quantitative measure of how effectively the data has been compressed. A higher compression ratio indicates more efficient compression, as it implies that the compressed data occupies less space compared to the original input data.

\textbf{Data Distortion.}
Data distortion is typically evaluated using the PSNR (Peak Signal-to-Noise Ratio), which is one of the most important metrics for assessing the quality of decompressed data resulting from lossy compression. PSNR is defined as follows:
\begin{equation}
  \text{PSNR} = 20 \log_{10} \texttt{vrange}(X) - 10 \log_{10} \texttt{mse}(X, X').
  \label{eq:psnr}
\end{equation}
Given input data $X$ and decompressed data $X'$, let $\texttt{vrange}$ represent the range of values within the input data array (the difference between the minimum and maximum values), $\texttt{mse}$ denotes the mean-squared error. Importantly, PSNR increases as $\texttt{mse}$ decreases, indicating an improvement in the quality of the decompressed data.

\subsection{Deep Neural Networks}
\label{section_dnn}
Deep neural networks (DNNs) are a class of artificial neural networks characterized by multiple layers of interconnected nodes, or neurons, in a hierarchical manner. Each layer in a DNN typically performs a nonlinear transformation of its inputs, allowing the network to extract higher-level features from the input data progressively. To train a DNN $f$, input data $x$ is fed forward through the model, generating predictions $\hat{y}=f(x;\Theta)$. The error between these predictions and the groundtruth $y$ is then calculated. To optimize the training objective $\min_{\Theta}\ell(f(x;\Theta), y)$, backpropagation computes gradients of the loss function $\ell(\cdot)$ with respect to the network's parameters $\Theta$, which are used to update them through optimization algorithms like stochastic gradient descent (SGD). 

The \textit{encoder-decoder} architecture is a special and fundamental network architecture of DNNs widely utilized in various image processing tasks. The \textit{encoder} transforms the input image into a fixed-dimensional latent representation by processing the input sequence through several layers of neural network operations, such as convolutional layers. This latent representation captures essential features of the input data, enabling effective information extraction. The \textit{decoder} then takes this latent representation as input and reconstructs the original data structure by generating an output through a series of decoding operations. To train the encoder-decoder model $f$ with parameters $\Theta$, we generally minimize the reconstruction error, i.e.,
\begin{equation}
\min_{\Theta}~ \ell(f(\hat{x}; \Theta), x), 
\label{eq:denoising}
\end{equation}
where $\hat{x}$ is the input image, $x$ is the target we want the model output, and the loss function $\ell(\cdot)$ can typically be the mean square error (MSE). Through this process, the encoder-decoder model learns a mapping from the input image and the output image. By its nature, this architecture facilitates tasks such as super-resolution \cite{wang2020deep}, image denoising \cite{fan2019brief}, image generation \cite{elasri2022image}, and image inpainting \cite{jam2021comprehensive}. Popular models such as Variational Autoencoder (VAE) \cite{kingma2013auto}, Generative Adversarial Network (GAN) \cite{goodfellow2014generative}, and U-Net \cite{ronneberger2015u} are based on encoder-decoder architectures.



\subsection{Problem Formulation}
\label{section_Problem}
Our research objective in this work is to improve the compression quality for error-bounded lossy compression significantly. The problem of our learning-based compression framework can be outlined as follows: How can we substantially enhance the fidelity of the reconstructed data (denoted by $X'$) from a structured mesh dataset (referred to as $X$) while adhering to strict user-defined error bound (i.e., $e$)? 

Drawing inspiration from the nature of image encoder-decoder architecture, i.e., the model learns a mapping from the input image to the output image, in our \proj~framework, we utilize encoder-decoder models to learn a mapping from the decompressed data by lossy compressors to the original scientific data. For example, we can regard our task as image denoising, as the slices in the decompressed data are noisy images, and the slices in the original data are the corresponding pure images. Then, we train the encoder-decoder model to denoise the input image by minimizing the reconstruction errors. To be specific, in our learning-based compression framework, we let the DNN-based models learn to transform the decompressed data to the original data and optimize the model parameters in the compression process. Mathematically, the optimization problem can be formally written as:
\begin{equation}\begin{aligned}
 \max_{\Theta} \ \ &\text{PSNR}(X, \mathcal{R}(X';\Theta)), \\
 \text{s.t.}\ \ \  &|x_i - x'_i| \leq e, \ \forall x_i \in X.
\end{aligned}
\label{eq:obj}
\end{equation}

\begin{figure*}[h]
  \centering
  \begin{minipage}[t!]{0.48\textwidth}
    \centering
    \includegraphics[width=0.825\linewidth]{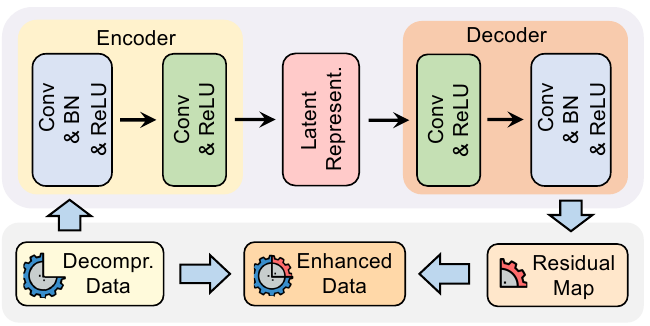}
    \caption{Illustration of \proj~learnable enhancer model design based on the encoder-decoder DNN architecture. `BN' represents the batch normalization layer.}
    \label{fig:residual-learning}
  \end{minipage}
  \hfill 
  \begin{minipage}[t!]{0.475\textwidth}
    \centering
    \includegraphics[width=\linewidth]{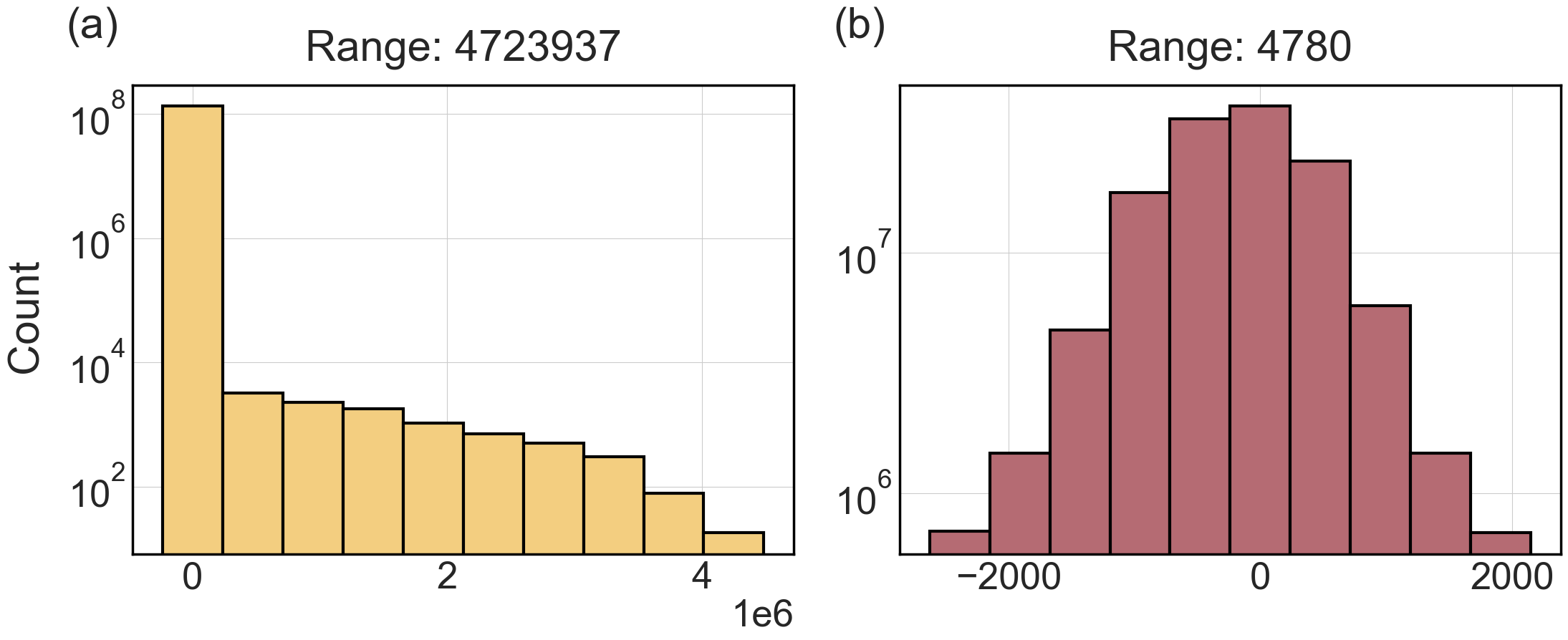}
    \caption{Distribution of (a) decompressed data and (b) residual data,  both generated by SZ3 compressor \cite{liang2022sz3} with a relative error bound of 5E-4 for Nyx dataset's Temperature field.}
    \label{fig:1gdistribution}
  \end{minipage}
\end{figure*}

Here, $\Theta$ are the parameters of the encoder-decoder DNN models $\mathcal{R}$ to be trained in the learning process within compression. The DNN models are considered \textbf{enhancers} to learn patterns from the difference between the original scientific input ($X$) and its decompressed representation ($X'$). By solving the problem formulated above, our \proj~can enjoy enhanced PSNR given a specified error bound. Alternatively, our \proj~can greatly improve the compression ratio with a fixed PSNR. The following sections will detail the specific implementation of \proj.

\section{\proj~--- Learn for Compression}

\label{overview}
In this section, we introduce the detailed design of our \proj. The objective of the proposed \proj~is to \ul{\textit{learn to improve the reconstruction quality with negligible overhead}} by leveraging multiple lightweight learnable enhancers to learn the mappings from the decompressed data to the original data in a group-wise manner.

\subsection{Design Overview}
The \proj~framework consists of the compression module and reconstruction module, which are illustrated in Figure \ref{fig:framework-compr} and Figure \ref{fig:framework-decompr}, respectively. 
 
In the compression module, the input scientific data, such as Nyx, HACC, and EXAFEL, are processed by a regular lossy compressor, e.g., SZ3, yielding compressed data. Since conventional lossy compressors need to know whether the decompressed value is error-bounded, the decompressed data are also generated in the compression process. Hence, \texttt{GWLZ} temporarily caches the decompressed data, between which and the original data, we calculate the residual map. Note that the residual map is of the same shape as the decompressed data and the original data. After that, \proj~partitions the decompressed data and the residual map into multiple groups according to value ranges, where their corresponding entries are in the same group. Then, for each partitioned data group, \proj~initializes a lightweight DNN model with an encoder-decoder structure, training the model to predict the residual map, which can be considered the compression error, as we feed the decompressed data into the DNN model. In the end, \proj~attaches the trained weights of the DNN models, whose size is negligible because of their lightweight nature, to the compressed data as the final compressed format. 

The reconstruction module is a reverse process against the compression module. In this module, \proj~first detaches the weights from the compressed format and uses them to initialize multiple DNN-based enhancer models. Then, \proj~decompresses the scientific data and partitions the decompressed data into multiple groups based on the same criterion in the compression module. Next, \proj~leverages the lightweight DNN models to predict the corresponding residual maps group by group. Finally, we add the residual map to the decompressed data and merge all the groups back to the original data format (illustrated in Figure \ref{fig:residual-learning}) as \textit{enhanced data} --- the output of our reconstruction module.


\subsection{Residual Learning: Improve Training Performance}
\label{section_residualmap}

The first challenge in the \proj~learning process is how to effectively learn to enhance the decompressed data to the original data. As shown in Table \ref{tab:nyx}, the gap between the maximal value and the minimal value in the Temperature field of the Nyx dataset is as large as $4.78\times 10^6$, which exacerbates the training instabilities caused by the gradient oscillation, leading to undesired training performance. 

Inspired by the DNN-based image denoising model, DnCNN \cite{zhang2017beyond}, which learns to predict the noise instead of the original pure image, we develop a similar convolutional neural network (CNN) model based on the encoder-decoder architecture and a residual learning strategy. To be specific, \proj~considers the slices of scientific data as single-channel images such that the decompressed data are noisy images. Our residual learning is to learn the residual information ($R=X-X'$) between the noisy decompressed data ($X'$) and the original data ($X$) slice by slice. The model design and the residual learning are demonstrated in Figure \ref{fig:residual-learning}. With the residual learning design, as shown in Figure \ref{fig:1gdistribution}(b), the magnitude range of the learning output is narrowed to thousands. Thus, the encoder-decoder enhancer model can enjoy stabilized learning and predict accurate residual information once well-trained, thereby enhancing the decompressed data closer to the original data. We have plotted the loss curves in Figure \ref{fig:group-wise-loss} in comparison with regular non-residual learning. It is seen that our residual learning strategy performs to lower the reconstruction error significantly.


Once the learning process is completed in the \proj~compression module, the DNN-based enhancer model predicts the residual map $\hat{R}$ from the decompressed data $X'$ by the lossy compressor in the \proj~reconstruction module. Subsequently, we add the predicted residual information to decompressed data, enhancing the reconstruction quality, i.e. $\hat{X}=X'+\hat{R}$.


\subsection{Group-wise Learning: Mitigate Biased Distribution}
\label{section_group-wise}


\begin{figure}[htbp]
  \centering
  \includegraphics[width=\linewidth]{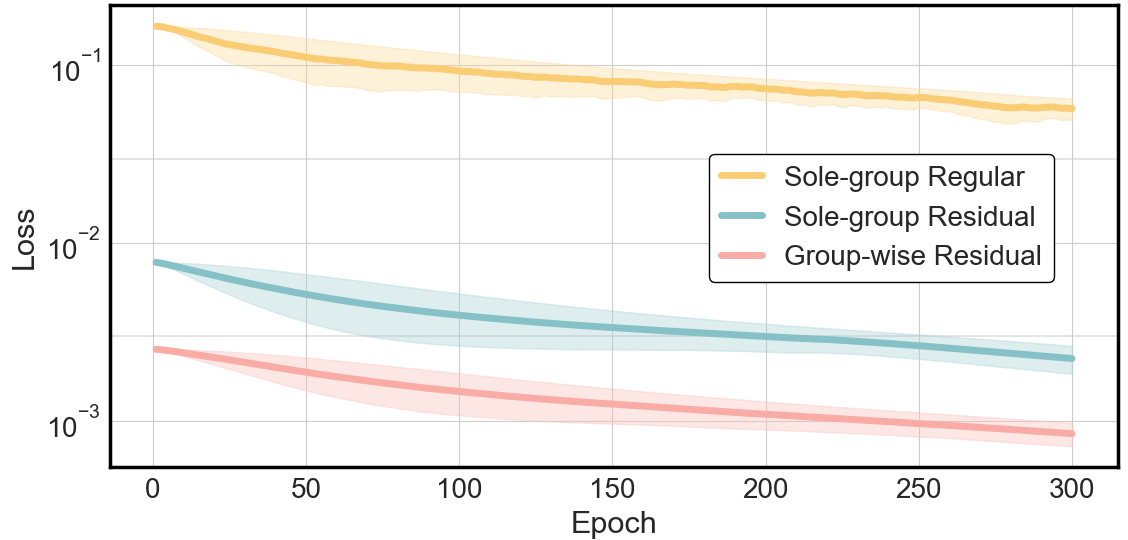}
  \vspace{-.3in}
  \caption{Comparison of loss curves during training (lower is better). `Sole-group Regular': learning to predict the original data. `Sole-group Residual': learning to predict the residual information. `Group-wise Residual': learning to predict the residual information in multiple groups separately.}
  \label{fig:group-wise-loss}
  \vspace{-2mm}
\end{figure}

With the residual learning introduced in the previous subsection, we have reduced the target learning range from huge values to ones that can be learned smoothly by an encoder-decoder DNN model. However, another significant challenge is that the learning in \proj~is from a \textit{biased skewed distribution to a normal distribution}. To be specific, the input data exhibits a skewed distribution, deviating significantly from a Gaussian-like pattern. This can be viewed as a `class imbalance' problem \cite{johnson2019survey}. The dominant class may disproportionately influence the model's gradient, increasing errors for the minority class and potentially hindering convergence.

Taking the Temperature field of the Nyx dataset as an example, Figure \ref{fig:1gdistribution} shows the distributions of the original scientific data and the corresponding residual map. The residual map represents the difference between the original data and the decompressed data generated by the SZ3 compressor \cite{liang2022sz3} with a relative error bound of 5E-4. The residual map exhibits a Gaussian-like distribution, while the original scientific data deviates significantly from a Gaussian-like pattern.
In the realm of deep learning, data distribution significantly impacts model performance. Inadequately distributed data can misdirect models, hindering convergence by leading them towards sub-optimal parameter configurations. As a consequence, enabling the learning from a skewed distribution to a normal distribution is very challenging.

Besides, the input data also has an extensive range between its minimum and maximum values (approximately 4.8M, as seen in Figure \ref{fig:1gdistribution}(a)). From the distribution, we can see that the majority of the values are concentrated in the former small range, while other values are scattered in the latter large range. After Min-Max scaling, a necessary preprocessing step, most data points shrink into an extremely narrow range. This phenomenon causes the vanishing of the majority of data, thereby hindering the DNN model from learning the most meaningful data and leading to unsatisfied performance improvement.

\begin{figure}[htbp]
  \centering
  \includegraphics[width=\linewidth]{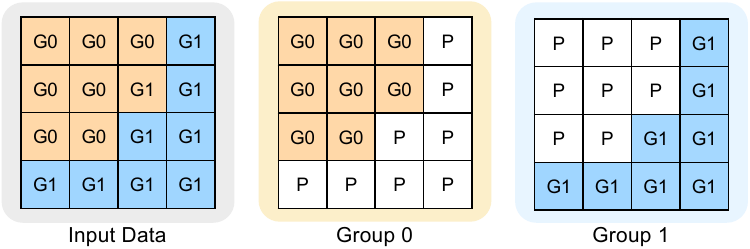}
  \vspace{-.3in}
  \caption{(Left): The input data is partitioned into two distinct groups. (Middle) and (Right): The DNN models are exclusively trained on the assigned group with other groups masked. `G': group; `P': placeholder.}
  \label{fig:groupmask}
  \vspace{-4mm}
\end{figure}

\begin{figure*}[t!]
  \centering
  \includegraphics[width=\linewidth]{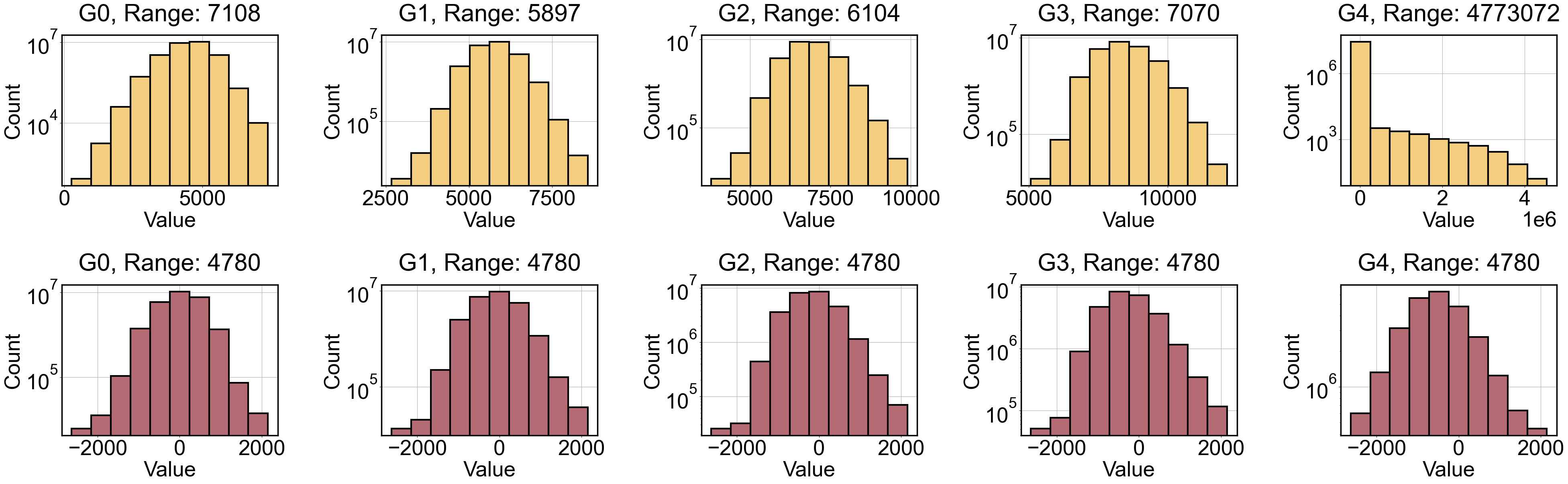}
  \caption{The data distribution for 5 groups of (Top) decompressed data and (Bottom) residual data, both generated by SZ3 compressor \cite{liang2022sz3} with a relative error bound of 5E-4 for the Temperature field from the Nyx dataset. `G': group.}
  \label{fig:5gdistribution}
  \vspace{-2mm}
\end{figure*}

Our motivation for the group-wise learning in \proj~is to narrow down the value range of the input data by dividing it into multiple groups based on value magnitude. Thus, we can train multiple DNN-based enhancer models separately on their corresponding data groups. In this way, the data to be fed into the DNN models are balanced and learnable after \proj~performs in-group data preprocessing, i.e., normalization, since each group has a small min-max gap. Surprisingly, with this value-based group strategy, the distributions of the majority of groups are normalized. We show the example distributions in Figure \ref{fig:5gdistribution} for a 5-group partition on the Temperature field of the Nyx dataset. It is seen that both the input data and the residual map exhibit Gaussian-like distribution in the first four groups. With the corrected data distribution, \proj~can enjoy a smooth and improved learning process, significantly enhancing the decompressed data quality in multiple groups separately by learning from a normal distribution to another normal distribution. Most importantly, thanks to the small sizes of groups, \proj~can initialize lightweight DNN models to learn very well within each group, thereby making the overall model overhead negligible. 

We have plotted the learning curves of our group-wise residual learning and sole-group residual learning in Figure \ref{fig:group-wise-loss}, which demonstrates that group-wise learning indeed reduces the learning error significantly. Specifically, both methods aim to learn the residual information between the decompressed data generated by SZ3 \cite{liang2022sz3} (with a relative error bound of 5E-3) and the original Temperature field data from the Nyx dataset. In the group-wise case, \proj~utilizes two DNN models, each of which is trained on a distinct partition of the data. The red curve represents the average loss across these two models.
While the sole-group case learns the residual information, the vast range between minimum and maximum values across the entire dataset, coupled with its undesirable distribution, leads to defective convergence performance. To understand the influence of DNN model size, we configured the sole-group case with a model possessing an equivalent total number of parameters as the two DNN models in the group-wise case. These results demonstrate the effectiveness of our group-wise learning strategy. As the number of groups increases, a significantly lower loss is attained, leading to improved reconstruction quality, which will be further explored in Section \ref{section_experimental-results}.

\textbf{Group and Training Strategy.} Figure \ref{fig:groupmask} provides a simplified example of a two-group partition. Given an input slice of the decompressed data, two partitioned groups are denoted as G0 and G1, respectively. The corresponding entries in the residual map are also assigned to their respective groups, ensuring a consistent pairing between input data and residual map. Then, G0 and G1 are assigned to two different DNN models independently. Entries from outside the group are assigned a placeholder `P' (assigned by zero in the actual training) with no semantic meaning to distinguish them from valid data during processing. In the back-propagation process, each DNN model focuses solely on its assigned group. Output entries corresponding to actual data entries within the designated group calculate and back-propagate gradients. Conversely, output entries corresponding to placeholder values are masked with zeros, effectively preventing them from participating in the gradient calculations.  This mechanism ensures that each DNN model learns exclusively from its assigned group, maintaining consistent pairing within the model structure. 

\section{Evaluation}

\begin{table}[t!]
    \vspace{+0mm}
    \caption{Properties of Nyx simulation data \cite{zhao2020sdrbench}.}
    \label{tab:nyx}
    \centering
    \begin{tabular}{cccccc}
    \toprule
    \textbf{Field} & \textbf{Min} & \textbf{Avg} & \textbf{Max} & \textbf{Range}\\
    \midrule
    Temperature  & 2281 &  8453  & 4783k  & 4780k  \\
    Dark Matter Density   &  0& 1 & 13779 &  13779 \\
    \bottomrule
    \end{tabular}
    \vspace{-0.4mm}
\end{table}

In this section, we detail the experimental setup employed to evaluate the performance of the newly proposed \proj~lossy compression algorithm. Following the setup description, we will present the experimental results accompanied by a thorough analysis.
\subsection{Experimental Setup}

\begin{table}[t!]
\vspace{+0.0mm}
\caption{\proj's PSNR enhancement results on two fields of the Nyx dataset. `REB' represents relative error bound.}
\label{tab:sz-gwlz-results}
\begin{tabular}{ccccc}
\toprule
\textbf{REB} &
  \textbf{\begin{tabular}[c]{@{}c@{}}PSNR\\ SZ3\end{tabular}} &
  \textbf{\begin{tabular}[c]{@{}c@{}}PSNR\\ \proj-20\end{tabular}} &
  \textbf{\begin{tabular}[c]{@{}c@{}}Improve $\uparrow$ \\ (\%)\end{tabular}} &
  \textbf{\begin{tabular}[c]{@{}c@{}} File Size\\ Overhead \\ \end{tabular}} \\
\midrule
\multicolumn{5}{c}{Temperature}         \\
\midrule
5E-3  & 60.7   & \textbf{73.0}   & +20.2   &0.071$\times$   \\
1E-3  & 72.8   & \textbf{80.8}   & +11.0   &0.014$\times$    \\
5E-4  & 77.6   & \textbf{83.8}   & +8.1    &0.007$\times$     \\
1E-4  & 88.0   & \textbf{91.9}   & +4.4    &0.0016$\times$     \\
5E-5  & 92.7   & \textbf{95.3}   & +2.8    &0.0009$\times$     \\
1E-5  & 105.1  & \textbf{106.7}  & +1.5    &0.0003$\times$     \\
\midrule
\multicolumn{5}{c}{Dark Matter Density} \\
\midrule
5E-3  & 72.4   & \textbf{77.6}   & +7.3    & 0.092$\times$  \\
1E-3  & 77.3   & \textbf{85.0}   & +9.9    & 0.016$\times$  \\
5E-4  & 80.7   & \textbf{88.4}   & +9.6    & 0.0077$\times$  \\
1E-4  & 89.6   & \textbf{97.7}   & +9.0    & 0.0019$\times$  \\
5E-5  & 93.4   & \textbf{102.1}  & +9.3    & 0.0010$\times$  \\
1E-5  & 105.0  & \textbf{112.3}  & +7.0    & 0.0003$\times$ \\
\bottomrule
\end{tabular}
\vspace{+0.4mm}
\end{table}

\textbf{Experimental Environment.} Our experiments are conducted on two servers, each equipped with two NVIDIA GEFORCE RTX 4090 GPUs, an Intel Xeon W5-3435X CPU (16 cores), and 512GB of DRAM.

\textbf{Datasets.}  We test two data fields, Temperature and Dark Matter Density, from the Nyx scientific dataset \cite{almgren2013nyx}. This dataset, originating from cosmological simulations, is frequently used to benchmark lossy compression algorithms \cite{tao2019optimizing,tao2019z, liu2021exploring, liu2023faz, liu2022dynamic, liu2023high, liu2023srn, zhao2020sdrbench, zhao2020significantly, zhao2021optimizing}. We use the Nyx sample dataset provided by SDRBench \cite{zhao2020sdrbench} to evaluate the performance of \proj.  

The dataset contains six fields (Temperature, Velocity\_X, Velocity\_Y, Velocity\_Z, Baryon Density, and Dark Matter Density), each with dimensions of $512 \times 512 \times 512$ in FP32 format, totaling 3.1GB in size. The statistical analysis \cite{zhao2020sdrbench} provides insights into data field characteristics, including descriptive statistics (minimum, average, maximum, range) and correlations. Notably, Temperature, Velocity\_X, Velocity\_Y, and Velocity\_Z are found to be highly correlated, whereas Baryon Density and Dark Matter Density shows low correlation. This study prioritizes experiments on Temperature and Dark Matter Density due to space constraints, with future work planned to investigate the remaining fields. 

\textbf{Learning Configurations.}
We employ multiple DNN models based on encoder-decoder architecture (Figure \ref{fig:residual-learning}) as enhancers in \proj. Each DNN model contains 9 channels, 2 convolutional layers, and approximately 200 parameters. The training process leverages 4 GPUs with a batch size of 10 over 300 epochs. We initialize the learning rate at 1E-3 and implement a decay of 0.005 every 30 epochs. Upon completion of training, all the weights of DNN models are attached to the final compressed data format with FP32 precision. 

\textbf{Compressor.}
As shown in Figures \ref{fig:framework-compr} and \ref{fig:framework-decompr}, our flexible \proj~framework can accommodate various lossy compressors. We select SZ compressors because of their superior compression quality, as demonstrated in existing literature \cite{zhao2020significantly}. Specifically, we choose SZ3 \cite{liang2022sz3}, a state-of-the-art one from the SZ compression family with a complex and multi-module design, to integrate into our \proj.


\subsection{Experimental Results}
\label{section_experimental-results}

Table \ref{tab:sz-gwlz-results} summarizes PSNR improvements across Temperature and Dark Matter Density fields under varying \textit{relative error bound} (REB) conditions. To demonstrate the advantages of group-wise learning, we conducted experiments with both a single-group baseline and a 20-group setup for each field under each REB. The 20-group cases consistently exhibit significant performance gains over the single-group baseline, validating the benefits of group-wise learning as outlined in Section \ref{section_group-wise}. 
For the Temperature field, under larger REB values (e.g., 5E-3), our method achieves a substantial PSNR improvement from 60.7 to 73.0 (approximately 20.2\%), while the baseline only reaches 63.8 (approximately 5.1\%). As REB decreases, the baseline's ability to improve PSNR diminishes, whereas our method maintains notable improvements. At REB 1E-5, our approach yields a 1.5\% PSNR increase, which is significant considering the difficulty of enhancing PSNR at high initial values.
Interestingly, the Dark Matter Density field exhibits consistent improvements around 10\% across different REB values. This difference from the Temperature field could be attributed to the distinct data distribution properties shown in Table \ref{tab:nyx}.

\begin{figure}[t!]
  \centering
  \includegraphics[width=\linewidth]{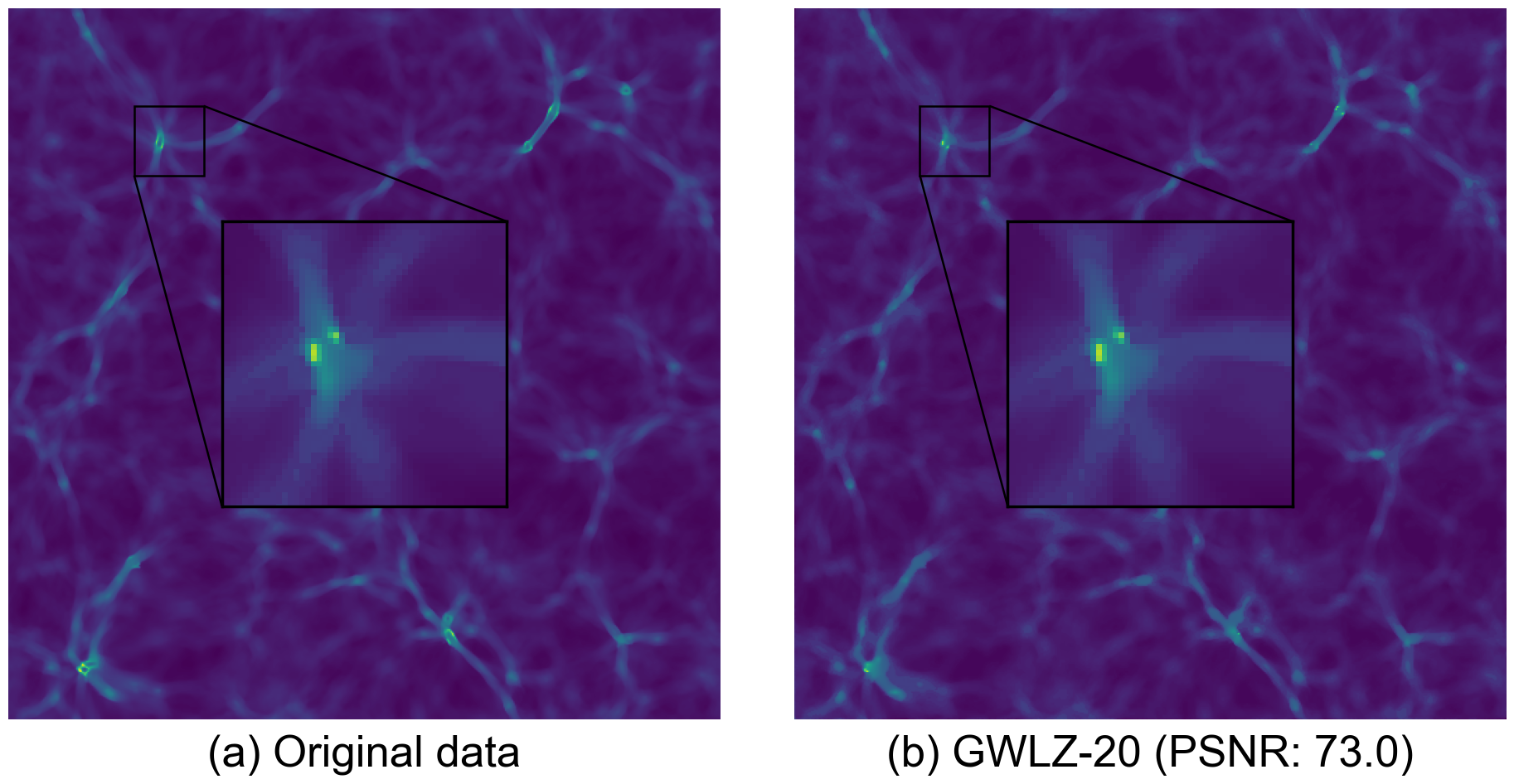}
  \vspace{-5mm}
  \caption{Visualization of enhanced snapshot data for the Nyx-Temperature.}
  \label{fig:visualizition}
  \vspace{0mm}
\end{figure}

Figure \ref{fig:visualizition} showcases \proj's high reconstruction quality by visualizing a slice from the enhanced Temperature field data (Nyx dataset). The zoomed region highlights \proj's ability to faithfully reconstruct local data patterns.  Indeed, the enhanced data exhibits a near-perfect visual match when compared to the original data.

These results demonstrate that our \proj~framework effectively improves the decompressed data of the two Nyx fields compressed with SZ3, yielding notable PSNR improvements. It's reasonable to expect similar results for other fields, datasets, and compressors, and we will do further experimentation to confirm the generalizability of \proj's benefits.

\begin{table}[t!]
\caption{PSNR performance of \proj~for multiple numbers of groups. \proj-$n$ denotes data are partitioned into $n$ groups.}
\label{tab:group-results}
\begin{tabular}{cccccc}
\toprule
\textbf{REB} &
  \textbf{\begin{tabular}[c]{@{}c@{}}PSNR\\ SZ3\end{tabular}} &
  \textbf{\begin{tabular}[c]{@{}c@{}}PSNR\\ \proj-1\end{tabular}} &
  \textbf{\begin{tabular}[c]{@{}c@{}}PSNR\\ \proj-5\end{tabular}} &
  \textbf{\begin{tabular}[c]{@{}c@{}}PSNR\\ \proj-10\end{tabular}} &
  \textbf{\begin{tabular}[c]{@{}c@{}}PSNR\\ \proj-20\end{tabular}} \\
\midrule
\multicolumn{6}{c}{Temperature}         \\
\midrule
5E-3  & 60.7   & 63.8   & 69.6   & 71.3  & \textbf{73.0}  \\
1E-3  & 72.8   & 73.1   & 76.8   & 78.8  & \textbf{80.8}  \\
5E-4  & 77.6   & 77.7   & 80.2   & 81.9  & \textbf{83.8}  \\
1E-4  & 88.0   & 88.1   & 89.3   & 90.3  & \textbf{91.9}  \\
5E-5  & 92.7   & 92.7   & 93.8   & 94.3  & \textbf{95.3}  \\
1E-5  & 105.1  & 105.1  & 105.7  & 106.0 & \textbf{106.7} \\
\midrule
\multicolumn{6}{c}{Dark Matter Density} \\
\midrule
5E-3  & 72.4   & 74.0   & 76.0   & 76.8  & \textbf{77.6}  \\
1E-3  & 77.3   & 79.8   & 82.7   & 84.0  & \textbf{85.0}  \\
5E-4  & 80.7   & 82.6   & 85.9   & 87.2  & \textbf{88.4}  \\
1E-4  & 89.6   & 90.4   & 94.0   & 95.6  & \textbf{97.7}  \\
5E-5  & 93.4   & 93.7   & 98.2   & 100.3 & \textbf{102.1} \\
1E-5  & 105.0  & 105.0  & 108.1  & 110.4 & \textbf{112.3} \\
\bottomrule
\end{tabular}
\vspace{-2mm}
\end{table}

\section{Discussions}
In this section, we analyze key findings from \proj~to understand the proposed \proj~further. We investigate the influence of the number of groups and model size, respectively.

\textbf{Number of Groups.} As discussed in Section \ref{section_group-wise}, increasing the number of groups facilitates a narrower range of values and promotes a desirable Gaussian-like distribution within each group. Table \ref{tab:group-results} confirms this trend, demonstrating that a larger number of groups correlates with higher PSNR improvements. This finding holds true across both the Temperature and Dark Matter Density fields, regardless of the REB value. Notably, the \proj-20 case consistently achieves the highest PSNR improvements. Additionally, the observed PSNR improvement exhibits a gradual increase as the number of groups goes up. This suggests a potential for continued quality enhancement with even more groups, although the existence of a definitive limit remains to be determined. While space limitations prevented testing configurations beyond 20 groups, our future research will conduct a comprehensive analysis to determine if a threshold exists for reconstruction quality improvement under an acceptable overhead as the number of groups increases.

\textbf{DNN Model Size.} As discussed in Section \ref{overview}, the DNN models in \proj~is trained to learn a mapping to enhance the reconstructed data. Generally, larger model size have the potential to `remember' more input data, leading to improved reconstruction quality. However, practical constraints limit the feasible size due to the infeasible overhead to the compressed file size.


Table \ref{tab:sz-gwlz-results} demonstrates the file size overhead incurred by the DNN-based enhancer models in \proj, which are appended to the compressed data to enable improved reconstruction. For instance, in the Temperature field with REB = 1E-5, the final compressed data with \proj~is only 1.0003$\times$ larger than the data compressed solely by SZ3. Values closer to 0 indicate better preservation of the original compression ratio. Notably, for both fields, smaller REB values correlate with file size overhead closer to 0. This aligns with the known difficulty of lossy compressors to achieve high compression ratios at smaller REBs \cite{tao2019z}.  Because DNN models in \proj~have a fixed size (e.g., 20 models with 200 parameters), their relative overhead decreases as the size of the compressed data increases at smaller REBs. This promising result suggests \proj~can significantly improve reconstruction quality with negligible overhead on compression efficiency.

\section{Conclusion}
In this paper we propose \proj, a novel learning-based lossy compression framework. By implementing multiple DNN models as lightweight enhancers, \proj~significantly enhances the decompressed data reconstruction quality, while maintaining remarkable compression efficiency. Experiments on diverse fields from the Nyx dataset demonstrate that \proj~achieves up to 20\% quality improvements with negligible storage overheads (as low as 0.0003$\times$). To the best of our knowledge, our \proj, is the first DNN-based learn-to-compress framework.


\bibliographystyle{ACM-Reference-Format}
\bibliography{ref}
\end{document}